\documentclass{stsci_report}

\usepackage{graphicx}
\usepackage{siunitx}
\usepackage{appendix}
\usepackage{tabularx}
\usepackage{url}
\usepackage{rotating}
\usepackage{tablefootnote}

\usepackage[roman]{parnotes}

\makeatletter
\def\parnoteclear{%
    \gdef\PN@text{}%
    \parnotereset
}
\makeatother

\usepackage[table]{xcolor}
\definecolor{lightgray}{gray}{0.85}
\definecolor{stblue}{HTML}{4C8BD8}
\usepackage{hyperref}
\usepackage{xcolor}
\usepackage{amssymb}
\usepackage{amsmath,siunitx}
\usepackage{array}
\usepackage{multirow}
\usepackage{enumitem}

\usepackage{wrapfig}

\usepackage{pdflscape}

\copyrighttext{Copyright\copyright\ \the\year\ The Association of Universities for Research in Astronomy, Inc. All Rights Reserved.}

\presubtitle{Instrument Science Report WFC3 2024-04}
\title{\textbf{WFC3/UVIS: External CTE Monitoring 2009 - 2024}}
\author{Benjamin Kuhn}
\date{April 30, 2024}

\begin{document}

\maketitle

\abstract{This report examines Charge Transfer Efficiency (CTE) flux losses in the Wide Field Camera 3 UVIS detector aboard the Hubble Space Telescope. Spanning approximately 14 years of observations from October 2009 to February 2024, the study analyzes CTE flux loss trends across various total background levels and source fluxes. In addition to analyzing the present state of CTE flux losses, we provide updated coefficients for the empirical model for point source photometry corrections in both non-CTE-corrected (FLT) and CTE-corrected (FLC) data. Between 2009 and 2023, the rate of CTE flux loss for a 500-2000 e$^-$ source, farthest from the readout, in FLT data with a 1-3 e$^-$/pix background, is measured to be $\sim$0.05 $\Delta$mag/2051 pix/year. The recommended minimum total background level to mitigate CTE losses remains at 20-25 e$^-$/pix. At that level, we find that 500-2000 e$^-$ sources, farthest from the readout, in 2024 FLT data can suffer $\sim$23$\%$ flux loss/2051 pix. The FLC data provide some relief, but we measure a $\sim$12$\%$ flux loss/2051 pix in 2024. There continues to be a slight over-correction in some FLC results that contain backgrounds above 40 e$^-$/pix. In 2024, 8000-20000 e$^-$ sources farthest from the readout in a 40, 60, or 90 e$^-$/pix background are over-corrected by $\sim$1, 2, and 3$\%$, respectively.}

\clearpage
\section{Introduction} \label{s:intro}
The charge transfer efficiency (CTE) of the Wide Field Camera 3 (WFC3) UVIS charge-coupled devices (CCDs) is a crucial detector characteristic that affects every exposure. CTE measures how well a charge packet is transferred from one pixel to the adjacent pixel during readout and depends on several factors: the distance (in pixels) from the readout, signal level of the source, total background level, and the amount of radiation damage. One of the perils of having CCDs in a space-based observatory such as the Hubble Space Telescope (HST) is the constant bombardment of cosmic rays and other high-energy radiation. 
\bigbreak

These high-energy radiation impacts have damaged some UVIS pixels, and ``charge traps" have formed. A trap can temporarily capture a fraction of a given charge packet as it transfers during readout and randomly releases it on timescales of microseconds to seconds \parencite{Baggett2011}. This delay in transferring the charge causes the detector to interpret the signal as originating upstream (in the anti-readout direction) of its true location and manifests in images as smeared trails extending from the core of sources. The smearing caused by the charge transfer inefficiency is a pathological issue that poses a problem for accurate photometry and astrometry \parencite{Anderson2021-13}. 
\bigbreak

With our current tools and understanding of CTE, the WFC3 team provides three ways to mitigate CTE flux losses post-observation: 
\vspace{0.1cm}
\begin{enumerate}[nolistsep]
    \item Use FLC files, which have been processed through the pixel-based CTE-correction algorithm included in \texttt{calwf3}\footnote{\texttt{calwf3} info: \href{https://hst-docs.stsci.edu/wfc3dhb/chapter-3-wfc3-data-calibration/3-1-the-calwf3-data-processing-pipeline}{https://hst-docs.stsci.edu/wfc3dhb/chapter-3-wfc3-data-calibration/3-1-the-calwf3-data-processing-pipeline }} \parencite{Anderson2021-9}.
    \item Use FLT files and apply the table-based corrections reported in WFC3 ISR 2021-13 \parencite{Anderson2021-13}.
    \item Use FLT or FLC files and apply the empirical model (shown in the \hyperlink{page.14}{Appendix}) using the coefficients closest to the exposure conditions to correct the fluxes. 
\end{enumerate}
\vspace{0.1cm}
Number 3 is the method we discuss in this report. The data presented here are input into the empirical model, and coefficients are re-derived for use by observers. 
\bigbreak

The WFC3 team has a yearly calibration program to measure the CTE flux loss using star cluster observations. The observing strategy involves taking two exposures separated by a one-chip dither, which causes all the sources on UVIS2 in the first image to land on UVIS1 in the second image. Due to the opposite readout directions of the CCDs, the stars close to the readout on UVIS2 are far from the readout in the subsequent image after the one-chip dither. Measuring the difference in the flux of the same star at varying distances from the readout enables us to characterize the photometric losses typical of charge transfer inefficiency. We measure the CTE flux losses as a function of the sky background by analyzing observations performed with different post-flash levels (see Section 2 and Table 1).
\bigbreak

Beginning in 2010, the CTE calibration programs observed 47 Tuc and NGC 6791 for their differing stellar densities because it was hypothesized that a high-density environment would cause a shielding effect (by pre-filling the charge traps) and interfere with measuring CTE flux losses. However, over time, this effect was not observed, and the measurements between the two targets are comparable within errors \parencite{Anderson2021-9}. Due to the sparse nature of NGC 6791, there are a limited number of stars in the field of view, which makes it difficult to measure CTE. For these reasons, CTE calibration programs stopped observing NGC 6791 and began observing Omega Centauri in 2020; Omega Centauri has a much more favorable density, providing 100s-1000s of stars per flux bin.
\bigbreak

In this report, we follow the standard analysis procedure, last detailed in \textcite{Kuhn2021}, and apply the model developed by \textcite{Noeske2012} to fit the data with a second-order polynomial constrained by source flux and epoch. The empirical model can be found in Appendix Section \hyperlink{page.14}{3}. We provide updated model coefficients that observers can use to correct non-CTE-corrected (FLT) and CTE-corrected (FLC) point source photometry. The bivariate empirical model presented here is completely separate from the pixel-based CTE-correction in the \texttt{calwf3} pipeline. The data in FLC files are not perfectly corrected and 100\% free from CTE losses. The results and model in this report are meant to be a supplemental correction applied to either FLC or FLT point source flux measurements. 
\bigbreak

For the latest CTE results and accompanying documentation, see our WFC3 CTE performance page\footnote{CTE performance page: \href{https://www.stsci.edu/hst/instrumentation/wfc3/performance/cte}{https://www.stsci.edu/hst/instrumentation/wfc3/performance/cte }}. This report presents losses for a subsample of the background levels. Observers interested in the coefficients and CTE flux loss plots for every background level should visit the CTE performance page.

\section{Data} \label{s:data}
The data consists of short (30 and 60 seconds) and long (348 and 420 seconds) exposures taken with the F502N filter. The first data for the UVIS external CTE program dates back to 2009. Since then, the WFC3 team has taken follow-up observations every cycle, giving us a baseline of 14 years. The programs are submitted without a specified HST orient, resulting in a different roll angle and, thus, a different set/placement of stars for each epoch. Starting in 2012, we have stimulated various background levels by post-flashing the detector at specific durations and current levels. With these differing exposure times and post-flash levels, we achieve total backgrounds between $\sim$0.1 - 120 e$^{-}$/pixel. While not every program has the same number of exposures or background levels, all programs since 2016 have observed two targets roughly twice a year at approximately seven different background levels. An overview of the program numbers, targets, approximate observation epochs, and the number of exposures used are listed in Table \ref{tab:1} in Section 1 of the Appendix.
\bigbreak 

The last Instrument Science Report to produce coefficients for the empirical model \parencite{Kuhn2021-6} included data up to July 2020 (ID \href{https://www.stsci.edu/cgi-bin/get-proposal-info?id=15721&submit=Go&observatory=HST}{15721}, PI Khandrika). This report includes additional data up to February 2024 (ID \href{https://www.stsci.edu/cgi-bin/get-proposal-info?id=17009&submit=Go&observatory=HST}{17009}, PI Kuhn). We use the same routine detailed in Section 4 of \textcite{Kuhn2021} to analyze the new exposures. This process includes aligning standard \texttt{calwf3} FLC files to custom reference catalogs with \texttt{Tweakreg} (\cite{drizzle}), propagating the resulting headerlet\footnote{Using headerlets: \href{https://hst-docs.stsci.edu/drizzpac/chapter-4-astrometric-information-in-the-header/4-6-using-headerlets}{https://hst-docs.stsci.edu/drizzpac/chapter-4-astrometric-information-in-the-header/4-6-using-headerlets }} to the corresponding FLT files, and performing background-subtracted aperture photometry (using a 3-pixel radius aperture) with \texttt{Photutils} (\cite{photutils}).

\section{Results} \label{s:results}
In addition to producing coefficients for the model, the main results of this analysis are separated into two types of plots, which we show in subsections 3.1 and 3.2. Each plot type has two panels that show CTE residual flux losses in non-CTE-corrected FLT files on the left and CTE-corrected FLC files on the right. Both plot types have y-axis units of $\Delta$ instrumental magnitude per 2051 pixels. Normalizing the results by 2051 represents the maximum number of row transfers a charge packet can experience during readout, which means \textbf{all the results presented are the maximum CTE residual flux losses for a given flux bin, with a certain total background level, at a specific epoch}.

\subsection{CTE Flux Losses as a Function of Flux Bin}

\begin{figure}[ht!]
  \centering
  \includegraphics[width=\linewidth]{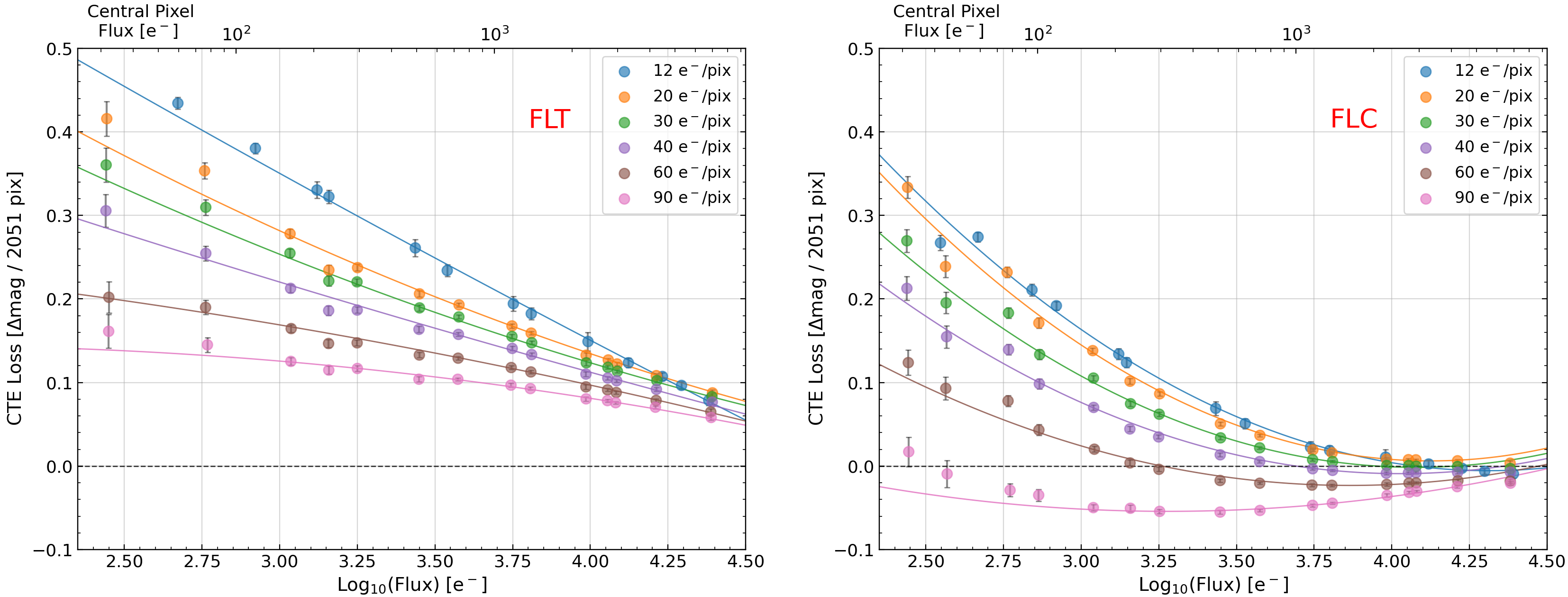}
    \caption{Residual CTE flux losses as a function of $log_{10}(flux)$ separated by approximate total background level in e$^-$/pix. Results for non-CTE-corrected FLT files and CTE-corrected FLC files are on the left and right, respectively (notated in red lettering at the top right of each panel). The upper x-axis on each panel shows the approximate central pixel flux in electrons. All the data come from February 16, 2024 observations of Omega Centauri.}\label{fig:omegacen}
\end{figure}

The first plot type can be seen in Figure \ref{fig:omegacen}, illustrating CTE residual flux loss as a function of $log_{10}$ of the flux (in e$^-$) in a background-subtracted 3-pixel radius aperture, with model fits over-plotted. The flux range on the x-axis spans $\sim$300-25000 e$^-$ (instrumental magnitude $\sim$ -6 through -11). The upper x-axis on each panel of Figure 1 shows the estimated central pixel flux in electrons. When a star is centered on a UVIS pixel in a visible wavelength filter, approximately 18\% of its flux will land in the central pixel \parencite{Anderson2021-9}. 
\bigbreak 

Figure 1 represents the updated state of CTE flux losses in UVIS, as they are our most recent observations from February 2024. Background levels between $\sim$ 12 and 90 e$^-$/pix are displayed, each denoted with a different color. The 12 e$^-$/pix background results, shown in blue, are from 30-second exposures. The remaining background levels are from 348-second exposures. The FLT results on the left side of Figure 1 show that CTE loss measurements decrease as the total background level increases. The sources in the 500-2000 e$^-$ flux bin ($log_{10}(flux)$ $\approx$ 3.03) experience $\sim$23$\%$ flux loss/2051 pix (0.28 $\Delta$mag/2051 pixel) in the 20 e$^-$/pix background and $\sim$11$\%$ (0.13 $\Delta$mag/2051 pixel) in the 90 e$^-$/pix background. The 500-2000 e$^-$ flux bin includes the most number of sources, varying between 10,919 stars in the 20 e$^-$/pix background and 11,253 stars in the 90 e$^-$/pix background level.
\bigbreak

The FLC results on the right side of Figure 1 also show that CTE improved with added post-flash. However, the 40, 60, and 90 e$^-$/pix background levels exhibit an over-correction in some, or most, flux bins. At the time of writing, the cause of the over-correction is unclear, but further investigations are planned. Sources with 500-2000 e$^-$ in the FLC images display a $\sim$12$\%$ flux loss (0.14 $\Delta$mag/2051 pixel) in the 20 e$^-$/pix background and a $\sim$ -5$\%$ (-0.05 $\Delta$mag/2051 pixel) in the 90 e$^-$/pix background. It should be noted that the pixel-based CTE correction that produces the FLC files purposefully restricts the correction on faint sources to limit the added noise-amplification level. The v2.0 CTE-correction algorithm measures the pixel-to-pixel noise in the background and, if it is less than 10 e$^-$, uses that as the noise-mitigation keyword. Then, the amount of reconstruction is dialed back whenever adjusting pixel values lower than the measured noise level, which results in fluctuations smaller than the background variations being mostly ignored \parencite{Anderson2021-9}.
\bigbreak

\begin{figure}[ht!]
  \centering
  \includegraphics[width=\linewidth]{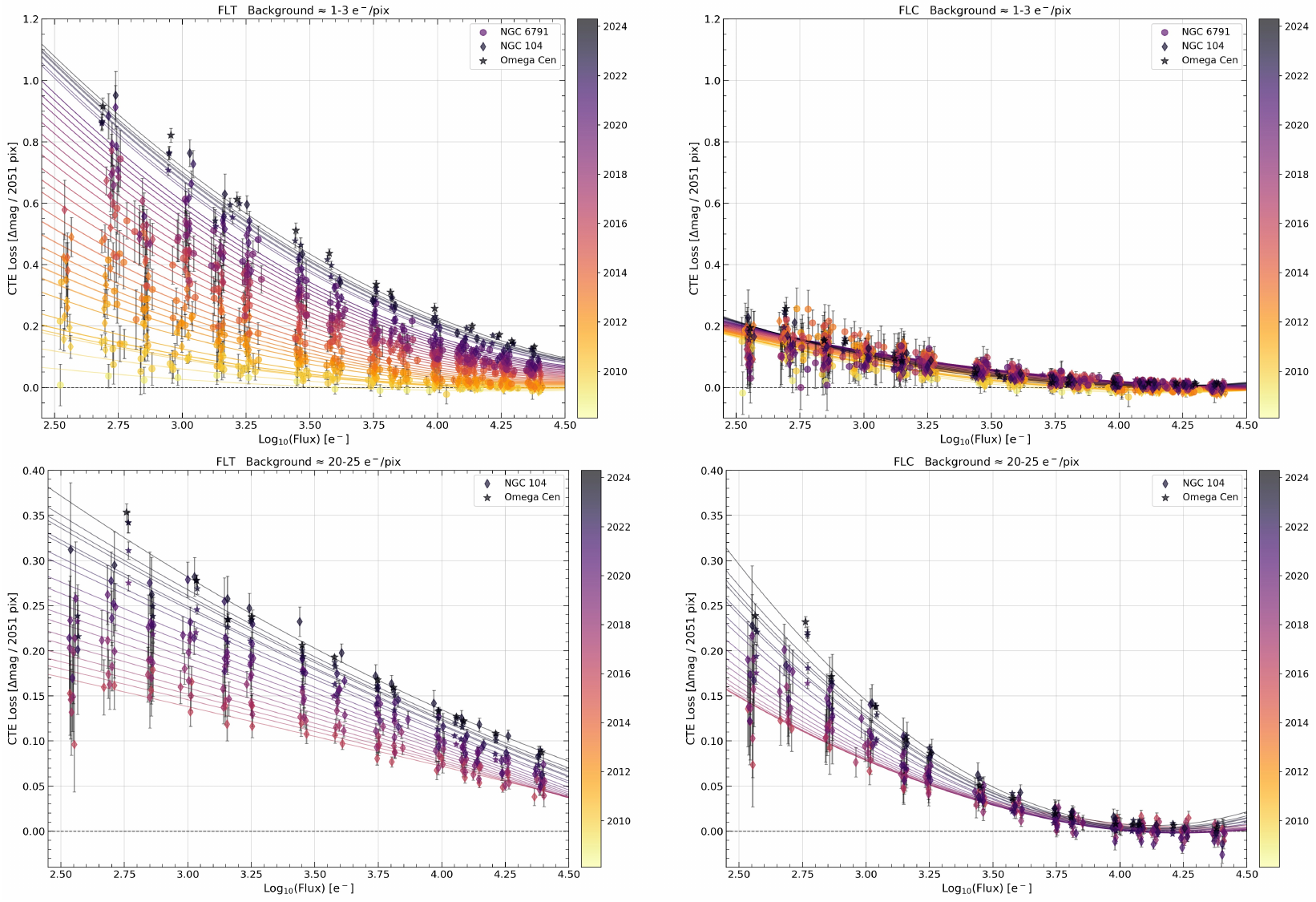}
    \caption{Residual CTE flux losses as a function of log$_{10}$ flux separated into two different background levels. The top row shows the evolution of losses in a $\sim$ 1-3 e$^-$/pix background between 2009-2024. The bottom row shows results in a $\sim$ 20-25 e$^-$/pix background between 2016-2024. Results for non-CTE-corrected FLT files and CTE-corrected FLC files are in the left and right columns, respectively.  The color coding corresponds to the observation date. Note that the y-axis range for the top and bottom rows differ. The top row spans -0.1 through 1.2, and the bottom ranges from -0.04 to 0.4.}
    \label{fig:cte_logf}
\end{figure}
In addition to Figure 1, we also provide Figure \ref{fig:cte_logf}, which again shows CTE residual flux loss as a function of $log_{10}$ of the flux (in e$^-$) in a background-subtracted 3-pixel radius aperture. However, rather than showing multiple background levels in a single panel, Figure 2 presents only two background levels, a low $\sim$ 1-3 e$^-$/pix background, and the currently recommended minimum background level, $\sim$ 20-25 e$^-$/pix. The color coding and attached color bars in Figure 2 illustrate the measurement's observation date. This type of plot allows us to see how CTE losses are getting larger with decreasing source signal levels and over time. The bottom row of Figure 2 (20-25 e$^-$/pix) has no yellow or orange data points because the baseline only extends back to 2016. 
\bigbreak 

The 1-3 e$^-$/pix background level seen in Figure 2 (and throughout the report) is achieved \textit{without} post-flash, meaning the background comes from only the natural sky (in F502N) and dark current. The 20-25 e$^-$/pix level is acquired \textit{with} post-flash; therefore, the background is composed of sky, dark current, and post-flash. Comparing the results on the top row with a low background to the results on the bottom with the minimum recommended total background shows how post-flashing to $\sim$20 e$^-$/pix reduces CTE flux losses. The ratio of CTE flux losses in the FLT data (top and bottom left panels of Figure 2) reveals that between 2016-2024, faint sources (2.5 $\leq$ $log_{10}(flux)$ $\leq$ 3.5) experience $\sim$ 2-3 times less flux loss in the 20-25 e$^-$/pix background.

\subsection{CTE Flux Losses as a Function of Date}

\begin{figure}[ht!]
  \centering
  \includegraphics[width=\linewidth]{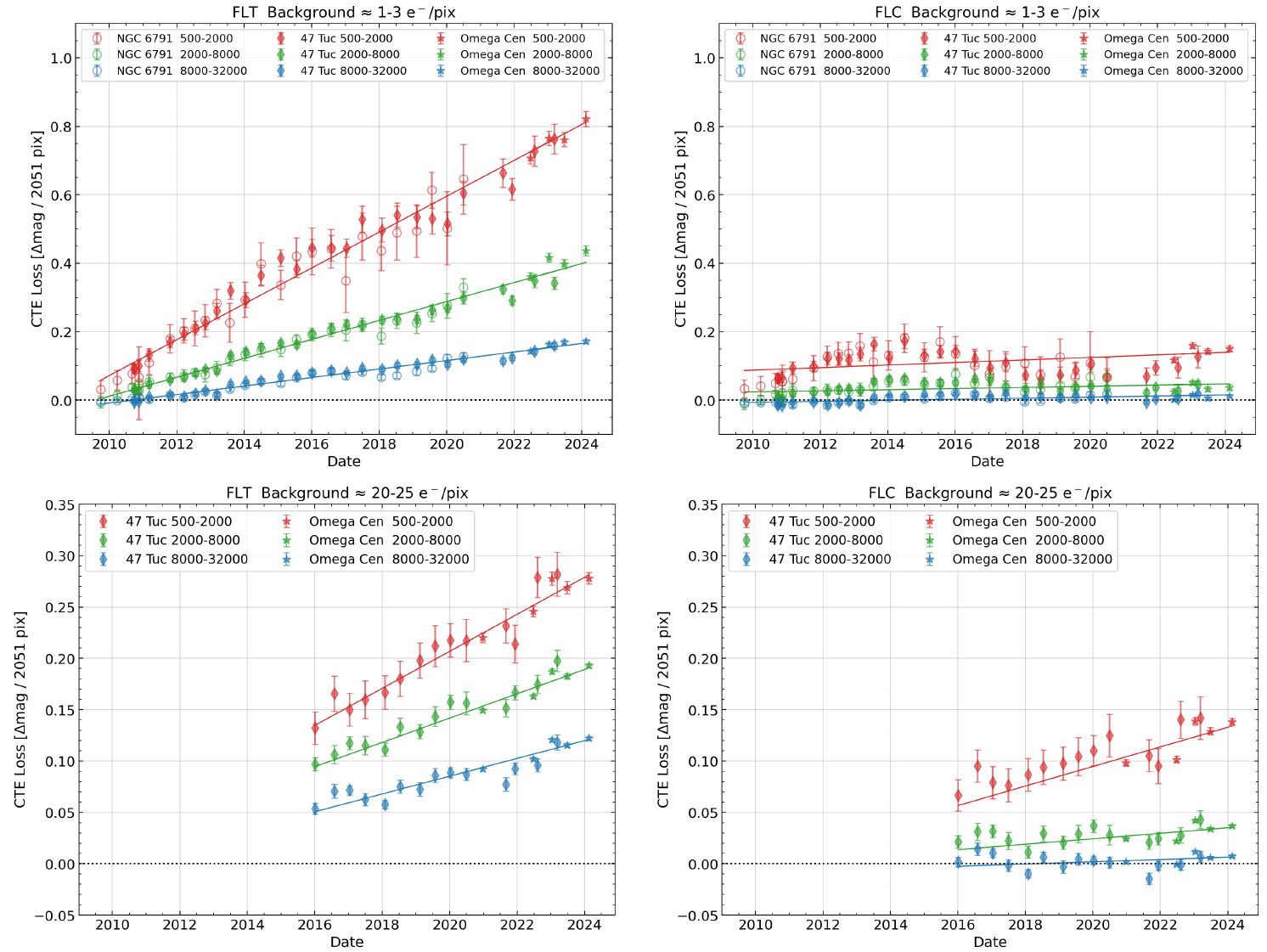}
    \caption{Residual CTE flux losses as a function of observation date separated into three main flux bins. As the legends show, the flux bins are 500-2000, 2000-8000, and 8000-32000 e$^-$. The top row shows the evolution of losses in a $\sim$ 1-3 e$^-$/pix background between 2009-2024. The bottom row shows results in a $\sim$ 20-25 e$^-$/pix background between 2016-2024. Results for non-CTE-corrected FLT files and CTE-corrected FLC files are in the left and right columns, respectively.}
    \label{fig:cte_date}
\end{figure}

The second plot type can be seen in Figure \ref{fig:cte_date}, where we show the evolution of CTE residual flux loss as a function of observation date for three main flux bins: 500-2000, 2000-8000, and 8000-32000 e$^-$ ($log_{10}$(flux) $\sim$ 3.0, 3.6, and 4.2). These flux bins correspond to the number of electrons measured in a background-subtracted 3-pixel radius aperture. Similar to Figure 2, in Figure 3, we display only the $\sim$ 1-3 and $\sim$ 20-25 e$^-$/pix background levels. Limiting the plot to three flux bins that span most of our full flux range provides a clearer view of how strongly the CTE losses depend on source brightness (in e$^-$) and observation date.
\bigbreak

Each flux bin has a weighted linear fit applied and over-plotted in the corresponding color. The red 500-2000 e$^-$ flux bin with a 1-3 e$^-$/pix background, in the upper left FLT panel of Figure 3, has a slope of $\sim$ 0.05 $\pm$ 0.001 $\Delta$mag/2051 pix/year or $\sim$ 3 $\pm$ 0.1$\%$ flux loss/2051 pix/year. The same flux bin in the panel below with a $\sim$ 20-25 e$^-$/pix background has a rate of change of $\sim$ 0.02 $\pm$ 0.001 $\Delta$mag/2051 pix/year or $\sim$ 2 $\pm$ 0.1$\%$ flux loss/2051 pix/year. 
\bigbreak

Looking on the right side of Figure 3, to the FLC results, the 500-2000 e$^-$ flux bin in the top right panel begins with an NGC 6791 measurement of $\sim$ 3 $\pm$ 2$\%$ flux loss/2051 pix at the end of 2009 that progresses to an Omega Cen measurement of $\sim$ 13 $\pm$ 1$\%$ flux loss/2051 pix in 2024. In the $\sim$ 20-25 e$^-$/pix background level, the same flux bin shows a $\sim$ 6 $\pm$ 1$\%$ flux loss/2051 pix in 2016 from 47 Tuc that climbs to an Omega Cen measurement of $\sim$ 12 $\pm$ 0.3$\%$ flux loss/2051 pix in 2024. The ``bump" or ``wave" feature seen (most dramatically) in the 500-2000 e$^-$ in the top right panel of Figure 3 is still not understood. 

\subsection{Tabulated CTE Flux Losses Results}
Tables \ref{tab:2} through \ref{tab:5} (in the Appendix) summarize select CTE flux loss results, listing the CTE residual flux loss slopes (like those seen in Figure 3) and the percent flux loss in the year containing the most recent data for nearly every background level the calibration program obtains. The slopes were calculated with a weighted least squares linear fit, and the percent flux losses were tabulated by evaluating the linear fit in the epoch containing the most recent data. 
\bigbreak 

The tables can be analyzed in two ways: 1) top to bottom for a single background level across multiple flux bins, or 2) left to right within a single flux bin across multiple background levels. For example, in Table 2, the 0.1-1 e$^-$/pix background results show sources with 500-2000, 2000-8000, and 8000-32000 e$^-$ in an aperture show slopes of $\sim$ 0.08, 0.04, and 0.01 $\Delta$mag/2051 pix/year (measured from late 2009 to early 2023). The corresponding percent flux losses in the 2023.19 epoch are listed to the right of the slopes at $\sim$ 67, 41, and 14$\%$ flux loss/2051 pix. An example of analyzing Table 2 within only the 500-2000 e$^-$ flux bin across multiple background levels shows 67, 53, 30, and 24$\%$ flux loss/2051 pix, in the 2023.19, 2024.13, or 2021.94 epoch, for backgrounds of $\sim$ 0.1-1, 1-3, 7-10, and 13-15 e$^-$/pix, respectively.    
\bigbreak

The newly derived coefficients, to be used with the model found in Appendix Section \hyperlink{page.14}{3}, are listed in Tables \ref{tab:6}-\ref{tab:9}. The coefficients are separated by total background level (e$^-$/pix) and file type (FLT Tables 6 \& 7 and FLC Tables 8 \& 9). Note that \textbf{the FLC coefficients should only be used to correct photometry in WFC3/UVIS FLC files with the v2.0 pixel-based CTE correction, first released in \texttt{calwf3 v3.6.0}.} Any FLC file retrieved from MAST\footnote{MAST: \href{https://mast.stsci.edu/portal/Mashup/Clients/Mast/Portal.html}{https://mast.stsci.edu/portal/Mashup/Clients/Mast/Portal.html}} since $\sim$ May 2021 will be  corrected with the v2.0 software. To verify the \texttt{calwf3} version and CTE-correction version, see the primary header keywords \texttt{CAL\_VER} and \texttt{CTE\_VER}.

\section{Conclusion} \label{s:conclusion}
As expected, CTE flux losses in the UVIS detector on board HST/WFC3 continue to increase over time. The number of high-energy radiation-induced charge traps grows year-to-year, which reduces the CTE of charge packets, especially those in pixels farthest from readout. Our yearly calibration efforts provide valuable insights into CTE flux loss trends and allow us to monitor the effectiveness of the post-flash and pixel-based CTE-correction. Swapping our observations of NGC 6791 for Omega Centauri in 2020 has drastically improved the number of sources obtained in each flux bin, ultimately giving our CTE flux loss measurements more confidence. 
\bigbreak 

With the baseline of our data now spanning between $\sim$ 8-14 years (depending on background level), we highlight the evolution of CTE flux loss patterns for various source signal levels observed in different total background levels. Our results illustrate the efficacy of post-flashing and why we still recommend a background level of $\sim$ 20 e$^-$/pix. While FLT data appear to benefit from post-flashing to higher background levels, our high background (high post-flash) FLC data show an apparent over-correction when the total background reaches $\gtrsim$ 40 e$^-$/pix. We will continue observing 47 Tuc and Omega Cen each cycle and plan to update the correction for high background at a future date.
\bigbreak 

To complement the external studies, we have recently added a supplemental internal CTE monitor (IDs \href{https://www.stsci.edu/cgi-bin/get-proposal-info?id=17259&submit=Go&observatory=HST}{17259} and \href{https://www.stsci.edu/cgi-bin/get-proposal-info?id=17356&submit=Go&observatory=HST}{17356}, PI Anderson). These programs aim to evaluate CTE losses and improve the pixel-based CTE correction by measuring the ``survival-to-readout" of electrons in warm pixels found in 500-second dark current exposures. To determine how CTE losses change with background level, post-flashing is employed to obtain 30 different levels ranging from 0 to 30000 e$^-$. Future results from these studies, along with follow-up external CTE monitor analyses like the one presented here, will provide evolving CTE loss measurements and may lead to potential new mitigation techniques and recommendations. 
\bigbreak 

Some of the key takeaways from this report include:
\begin{itemize}
    \item CTE flux losses will always increase over time; in addition, the level of the losses at any given observation date depends on the source flux, background level, and the distance in pixels from the readout. There is no perfect correction for CTE losses, and losses will affect every UVIS image (FLT and FLC) at some level. 
    \item The main results from this work are the updated coefficients listed in \hyperref[tab:6]{Tables 6-9} in the Appendix to be used with the CTE \hyperlink{page.14}{empirical correction model. }
    \item CTE residual flux loss results seen in Figures \ref{fig:omegacen}-\ref{fig:cte_date}, measured in units of $\Delta$mag/2051 pix, represent the maximum losses (i.e.\ ``worst-case scenario") for a given flux bin, with a certain total background level, at a specific epoch in time.
    \item \hyperref[tab:2]{Tables 2-5}  (in the Appendix) show the tabulated CTE residual flux loss slopes and percent flux loss measurements for the main flux bins and various background levels.
    \item In the most extreme case (not used for science data), faint sources with 250-500 e$^-$ in a 3-pixel radius aperture, farthest from readout, with a low 0.1-1 e$^-$/pix background level can experience $\sim$ 83$\%$ flux loss in 2023 FLT data. In contrast, the same faint sources, farthest from readout, with the moderate recommended $\sim$ 20-25 e$^-$/pix background, lose only $\sim$ 26$\%$ in 2023 FLT data.
    \item Observers interested in the plots and coefficients for every available background level ($\sim$ 0.1 - 120 e$^-$/pixel) should see our \href{http://www.stsci.edu/hst/instrumentation/wfc3/performance/cte}{WFC3 CTE performance page}.
\end{itemize}


\section*{Acknowledgements}
The author thanks Sylvia Baggett for her continued guidance throughout this project and for thoroughly reviewing the report. Additionally, the author graciously recognizes Jay Anderson, Annalisa Calamida, Mariarosa Marinelli, and WFC3 ISR editor Joel Green for reviewing the report and providing useful feedback that strengthened the ISR.

\newpage
\printbibliography

\newpage
\section*{Appendix}
\setcounter{section}{0}
\subsubsection*{  1 \quad Data Table Referenced in Section \ref{s:data} Above}\label{appendix_1}
\vspace{-.5cm}
 \begin{table}[hbt!]
 \begin{center}
 \caption{UVIS External CTE Observations Used in This Report\label{tab:1}}
    \vspace{.15cm}
    \begin{tabularx}{0.97\linewidth}{|c|c|c|c|c|}
    \hline \hline  
    \rowcolor{white}
	 Program ID & Principal Investigator &  Targets &  Observation Epochs & Number of  \\
    \rowcolor{white}
                 &                        &           &  (MM/YYYY) &          Exposures \\
    \hline 
    \rowcolor{lightgray}
	11924  &  Kozhurina-Platais  &  NGC 6791 &  10/2009, 03/2010,  & 12\\
    \rowcolor{lightgray}
            &                     &           &   09/2010 &          \\
	\hline
	\rowcolor{white}
	12348  &  Baggett  &  47 Tuc. &  09/2010  & 8 \\
	\hline
	\rowcolor{lightgray}
	12379  &  Noeske  & 47 Tuc.,  &  11/2010, 03/2011  & 15 \\
    \rowcolor{lightgray}
           &          & NGC 6791  &                  &   \\ 
	\hline
	\rowcolor{white}
	12692  &  Noeske  &  47 Tuc.,   &  10/2011, 03/2012, & 18    \\
    \rowcolor{white}
           &          &  NGC 6791   &  07/2012          &         \\            
	\hline
	\rowcolor{lightgray}
	13083  &  Noeske  &  47 Tuc.,   &  11/2012, 03/2013, & 60     \\
    \rowcolor{lightgray}
          &           &  NGC 6791   &  07/2013, 08/2013  &       \\
	\hline
	\rowcolor{white}
	13566  &  Noeske  &  47 Tuc.,  &  01/2014, 07/2014 &  40   \\
    \rowcolor{white}
          &           &  NGC 6791   &                &            \\
	\hline
	\rowcolor{lightgray}
	14012  &  Gosmeyer  &  47 Tuc.,  &  01/2015, 02/2015, &   40   \\
    \rowcolor{lightgray}
          &           &  NGC 6791   &   07/2015       &           \\
	\hline
	\rowcolor{white}
	14378  &  Mack  &  47 Tuc.,   &  01/2016, 07/2016,  &    56  \\
    \rowcolor{white}
          &           &  NGC 6791   & 08/2016       &           \\
	\hline
	\rowcolor{lightgray}
	14541  &  Mack  &  47 Tuc.,   &  01/2017, 07/2017 &   56 \\
    \rowcolor{lightgray}
          &           &  NGC 6791   &             &               \\
	\hline
	\rowcolor{white}
	14990  &  Fowler  &  47 Tuc.,   &  01/2018, 07/2018 &   56 \\
    \rowcolor{white}
          &           &  NGC 6791   &                  &          \\
	\hline
	\rowcolor{lightgray}
	15576  &  Kurtz  & 47 Tuc.,   &  02/2019, 07/2019 &    56 \\
    \rowcolor{lightgray}
          &           &  NGC 6791   &                 &           \\
	\hline
	\rowcolor{white}
	15721 &  Khandrika  & 47 Tuc.,   &  01/2020, 06/2020, &   56   \\
    \rowcolor{white}
          &           &  NGC 6791   &   07/2020         &                \\
    \hline
    \rowcolor{lightgray}
	16401 &  Kuhn  & 47 Tuc.,   &  12/2020, 09/2021,   &  56  \\
    \rowcolor{lightgray}
        &           &  Omega Cen. & 12/2021       &                     \\
    \hline
    \rowcolor{white}
	16573 &  Kuhn  &  47 Tuc.,   &  06/2022, 08/2022,  &    52 \\
    \rowcolor{white}
        &           &  Omega Cen. & 01/2023, 03/2023  &          \\
    \hline
    \rowcolor{lightgray}
	17009 &  Kuhn  &  47 Tuc.,   &  06/2023, 09/2023 &   42 \\
    \rowcolor{lightgray}
        &           &  Omega Cen. &   02/2024          &            \\
    \hline
\end{tabularx}
\end{center}
\hyperlink{page.3}{ Jump Back to Section 2: Data }
\end{table}


\subsubsection*{  2 \quad CTE Flux Loss Results Referenced in Section \ref{s:results} Above}\label{appendix_2}
\begin{table}[hbt!]
 \begin{center}
 \caption{CTE Flux Loss Evolution and Percent Flux Loss in FLT files\label{tab:2}}
    \vspace{.15cm}
\begin{tabularx}{\linewidth}{|c||X|X|X|X|X|X|X|X|}
\hline \hline  
\rowcolor{white}
Background [e$^-$/pix]  & \multicolumn{2}{c|}{0.1-0.5} & \multicolumn{2}{c|} {1-3} & \multicolumn{2}{c|}{7-10} & \multicolumn{2}{c|} {13-15} \\\hline
\rowcolor{lightgray}
\parnotereset Flux Bin [e$^-$] & CTE loss rate\emph{\parnote{$\Delta$mag/2051 pix/yr}} & CTE flux loss\emph{\parnote{$\frac{\text{\% flux loss}}{\text{2051 pix}}$ in 2023.19}}   & CTE loss rate$^i$   &  CTE flux loss\emph{\parnote{$\frac{\text{\% flux loss}}{\text{2051 pix}}$ in 2024.13}}  & CTE loss rate$^i$  &  CTE flux loss\emph{\parnote{$\frac{\text{\% flux loss}}{\text{2051 pix}}$ in 2021.94}}  & CTE loss rate$^i$  &  CTE flux loss$^{iv}$ \\\hline
\rowcolor{white}
500-2000  &  0.08    &  67$\%$    &  0.05  &   53$\%$   & 0.03    &  30$\%$ &   0.02    &  24$\%$   \\\hline
\rowcolor{lightgray}
2000-8000  &  0.04    &  41$\%$   & 0.03  &  31$\%$    & 0.02    &  18$\%$ &    0.02     &  16$\%$   \\\hline
\rowcolor{white}
8000-32000  &  0.01   &  14$\%$   &  0.01  & 14$\%$   &  0.01   &   10$\%$  &   0.01      &  9$\%$ \\\hline
\end{tabularx}
    \parnotes
 \end{center}
    \vspace{-.25cm}
    Table 2: CTE residual flux loss evolution rate and percent flux loss results for four background levels in FLT data.  The CTE flux loss rates and percentages listed are for sources furthest from the readout amplifiers (maximum number of transfers) in the 2021, 2023, or 2024 epoch. The CTE flux loss evolution for the 0.1-0.5 and 1-3 e$^-$/pix background has a baseline of 13-14 years. The CTE flux loss evolution for data with a 7-10 and 13-15 e$^-$/pix background level has a baseline of 9 years (2012-2021). The units for each column are noted with the $i$, $ii$, $iii$, and $iv$ superscripts and listed under the table with dates shown in decimal year.
\end{table}
\vspace{-0.5cm}
\begin{table}[hbt!]
 \begin{center}
 \caption{CTE Flux Loss Evolution and Percent Flux Loss in FLC files\label{tab:3}}
    \vspace{.15cm}
\begin{tabularx}{\linewidth}{|c||X|X|X|X|X|X|X|X|}
\hline \hline  
\rowcolor{white}
Background [e$^-$/pix]  & \multicolumn{2}{c|}{0.1-0.5} & \multicolumn{2}{c|} {1-3} & \multicolumn{2}{c|}{7-10} & \multicolumn{2}{c|} {13-15} \\
\hline
\rowcolor{lightgray}
\parnotereset Flux Bin [e$^-$] & CTE loss rate\emph{\parnote{$\Delta$mag/2051 pix/yr}} & CTE flux loss\emph{\parnote{$\frac{\text{\% flux loss}}{\text{2051 pix}}$ in 2023.19}}   & CTE loss rate$^i$   &  CTE flux loss\emph{\parnote{$\frac{\text{\% flux loss}}{\text{2051 pix}}$ in 2024.13}}  & CTE loss rate$^i$  &  CTE flux loss\emph{\parnote{$\frac{\text{\% flux loss}}{\text{2051 pix}}$ in 2021.94}}  & CTE loss rate$^i$  &  CTE flux loss$^{iv}$ \\
\hline
\rowcolor{white}
500-2000  &  0.01              &  24$\%$             & \small{$<$}0.01    &   12$\%$   & 0.01             &  18$\%$ &   0.01    &  14$\%$   \\
\hline
\rowcolor{lightgray}
2000-8000  &  \small{$<$}0.01   &  6$\%$             & \small{$<$}0.01     &  4$\%$    & \small{$<$}0.01    &  6$\%$ &    \small{$<$}0.01     &  4$\%$   \\
\hline
\rowcolor{white}
8000-32000  & \small{$<<$}0.01  &  \small{$\sim$}0$\%$  &  \small{$<$}0.01  & 1$\%$   &  \small{$<$}0.01            &   0.8$\%$  &   \small{$<<$}0.01  &  0.4$\%$ \\\hline
\end{tabularx}
    \parnotes
 \end{center}
    \vspace{-.25cm}
    Table 3: CTE residual flux loss evolution and percent flux loss results for four background levels in pixel-based CTE-correction applied FLC data.  The CTE flux loss rates and percentages listed are for sources furthest from the readout amplifiers (maximum number of transfers) in the 2021, 2023, or 2024 epoch. The CTE flux loss evolution for the 0.1-0.5 and 1-3 e$^-$/pix background has a baseline of 13-14 years. The CTE flux loss evolution for data with a 7-10 and 13-15 e$^-$/pix background level has a baseline of 9 years (2012-2021). The units for each column are noted with the $i$, $ii$, $iii$, and $iv$ superscripts and listed under the table with dates shown in decimal year. \\
    \\
    \hyperlink{page.7}{ Jump Back to Figure 3 }
\end{table}
\begin{table}[hbt!]
 \begin{center}
 \caption{CTE Flux Loss Evolution and Percent Flux Loss in FLT files\label{tab:4}}
    \vspace{.15cm}
    \begin{tabularx}{\linewidth}{|c||X|X|X|X|X|X|X|X|}
    \hline \hline  
    \rowcolor{white}
	 Background [e$^-$/pix]  & \multicolumn{2}{c|}{20-25} & \multicolumn{2}{c|} {30-35} & \multicolumn{2}{c|}{40-45} & \multicolumn{2}{c|} {60-65} \\\hline
	\rowcolor{lightgray}
	  \parnotereset Flux Bin [e$^-$] & CTE loss rate\emph{\parnote{$\Delta$mag/2051 pix/yr}} & CTE flux loss\emph{\parnote{$\%$ flux loss/2051 pix in 2024.13}}   & CTE loss rate$^i$   &  CTE flux loss$^{ii}$  & CTE loss rate$^i$  &  CTE flux loss$^{ii}$  & CTE loss rate$^i$  &  CTE flux loss$^{ii}$ \\\hline
	\rowcolor{white}
	500-2000  &  0.02             & 23$\%$     &  0.02           &   21$\%$  &  0.02     &  18$\%$      & 0.01   & 14$\%$   \\\hline
	\rowcolor{lightgray}
	2000-8000  &  0.01            &  16$\%$    & 0.01              &  15$\%$    & 0.01    &  14$\%$        & 0.01   &  11$\%$  \\\hline
	\rowcolor{white}
	8000-32000  &  \small{$<$}0.01  &  11$\%$   & \small{$<$}0.01  &  10$\%$    & \small{$<$}0.01 & 9$\%$   & \small{$<$}0.01  & 8$\%$ \\\hline
    \end{tabularx}
    \parnotes
 \end{center}
    \vspace{-.25cm}
    Table 4: CTE residual flux loss evolution and percent flux loss results for four background levels in FLT data. The CTE flux loss rates and percentages listed are for sources furthest from the readout amplifiers (maximum number of transfers) in the 2024 epoch. The CTE flux loss evolution for data with a 20-25 and 30-35 e$^-$/pix background level has a baseline of 8 years (2016-2024). The CTE flux loss evolution for the 40-45 and 60-65 e$^-$/pix background data has a baseline of 12 years (2012-2024). The units for each column are noted with the $i$ and $ii$ superscripts and listed under the table with the date shown in decimal year.
\end{table}
\vspace{-0.5cm}
\begin{table}[hbt!]
 \begin{center}
 \caption{CTE Flux Loss Evolution and Percent Flux Loss in FLC files\label{tab:5}}
    \vspace{.15cm}
    \begin{tabularx}{\linewidth}{|c||X|X|X|X|X|X|X|X|}
    \hline \hline  
    \rowcolor{white}
	 Background \small{[e$^-$/pix]} & \multicolumn{2}{c|}{20-25} & \multicolumn{2}{c|} {30-35} & \multicolumn{2}{c|}{40-45} & \multicolumn{2}{c|} {60-65} \\\hline
	\rowcolor{lightgray}
	  \parnotereset Flux Bin [e$^-$] & CTE loss rate\emph{\parnote{$\Delta$mag/2051 pix/yr}} & CTE flux loss\emph{\parnote{$\%$ flux loss/2051 pix in 2024.13}}   & CTE loss rate$^i$   &  CTE flux loss$^{ii}$  & CTE loss rate$^i$  &  CTE flux loss$^{ii}$  & CTE loss rate$^i$  &  CTE flux loss$^{ii}$ \\\hline
	\rowcolor{white}
	500-2000     &  0.01             & 12$\%$    &  0.01         &  9$\%$            &  \small{$<$}0.01     &  7$\%$            & \small{$<<$}0.01         & 2$\%$   \\\hline
	\rowcolor{lightgray}
	2000-8000    &  \small{$<$}0.01   &  3$\%$    & \small{$<<$}0.01    &  2$\%$         & \small{$<<$}0.01    &  0.7$\%$          & \small{$<$-0.01}   &  \small{$\sim$}0$\%$  \\\hline
	\rowcolor{white}
	8000-32000   &   \small{$<$}0.01  &  0.6$\%$  & \small{$<<$}0.01  &  \small{$\sim$}0$\%$  & \small{$<<$}0.01   & \small{$\sim$}0$\%$   & \small{$<<$-0.01}  & \small{$\sim$}0$\%$ \\\hline
    \end{tabularx}
    \parnotes
 \end{center}
    \vspace{-.25cm}
    Table 5: CTE residual flux loss evolution and percent flux loss results for four background levels in pixel-based CTE-correction applied FLC data. The CTE flux loss rates and percentages listed are for sources furthest from the readout amplifiers (maximum number of transfers) in the 2024 epoch. The CTE flux loss evolution for data with a 20-25 and 30-35 e$^-$/pix background level has a baseline of 8 years (2016-2024). The CTE flux loss evolution for the 40-45 and 60-65 e$^-$/pix background data has a baseline of 12 years (2012-2024). The units for each column are noted with the $i$ and $ii$ superscripts and listed under the table with the date shown in decimal year.\\
    \\
    \hyperlink{page.7}{ Jump Back to Figure 3 }
\end{table}
\clearpage

\subsubsection*{  3 \quad Empirical Photometric Correction Model}

As presented in \textcite{Noeske2012}, the bivariate polynomial used to estimate CTE-induced flux loss is defined in terms of $S$: 
\begin{equation}
    S = \sum_{i,j=0}^{2} c_{ij} d^{i} f^{j}
\end{equation}
where $c_{ij}$ are the polynomial coefficients, $f$ is the $log_{10}$(Flux) of the source, and the observation date is: \\

\vspace{-.7cm}
\begin{center}
$d = MJD - 55400$
\end{center}
\vspace{-.4cm}

Expanding equation 1 then gives:
\begin{equation}\label{eq2}
    S = c_{00} + c_{01}f + c_{02}f^{2} + d (c_{10} + c_{11}f + c_{12}f^{2}) + d^{2}(c_{20} + c_{21}f + c_{22}f^{2})
\end{equation}
Using the coefficients provided below in Tables \ref{tab:6} - \ref{tab:9} and solving for $S$ in Equation 2, observers can estimate the CTE-corrected magnitude of their point-source photometry with: 
\begin{equation}
    m_{corr} = m_{0} - S \frac{Y}{2051}
\end{equation}
where $m_0$ is the uncorrected magnitude, and $Y$ is the number of detector rows the source is away from the amplifier. The coefficients chosen for the model should correspond to the conditions that best match the observations attempting to be corrected. The coefficients are separated by total background level (e$^-$/pix) and file type (FLT and FLC). The FLC coefficients below should only be used to correct photometry in UVIS FLC files with the v2.0 pixel-based CTE correction, first released in \texttt{calwf3} v3.6.0. Any FLC file retrieved from MAST\footnote{MAST: \href{https://mast.stsci.edu/portal/Mashup/Clients/Mast/Portal.html}{https://mast.stsci.edu/portal/Mashup/Clients/Mast/Portal.html}} since $\sim$ May 2021 will have been corrected by the v2.0 software. The coefficients are also available to download as text files from the WFC3 CTE performance page\footnote{CTE Performance Page: \href{https://www.stsci.edu/hst/instrumentation/wfc3/performance/cte}{https://www.stsci.edu/hst/instrumentation/wfc3/performance/cte}}.    
\newline 
\newline
\hyperlink{page.2}{ Jump Back to Section 1: Introduction }
\newline
\newline
\hyperlink{page.8}{ Jump Back to Section 4: Conclusion }
\clearpage
\subsubsection*{  4 \quad Coefficients to Use with the Model Referenced in Appendix Section 3}\label{appendix_3}
\begin{table}[h!]
 \begin{center}
 \caption{Empirical CTE model coefficients for non-CTE-corrected FLT data \label{tab:6}}
    \vspace{.15cm}
    \begin{tabularx}{\linewidth}{|X|c|c|c|c|}
    \hline \hline  
	Filter, Exposure Time & F502N, Short & F502N, Short & F502N, Long & F502N, Long \\
	 Background [e$^-$/pix]  & $\sim$ 0.1-0.5  & $\sim$ 12-12.5  & $\sim$ 1-3  & $\sim$ 7-10  \\
    \hline 
    \rowcolor{lightgray}
	C$_{00}$  & 7.36e-01  &  -3.03e-01 &  8.59e-01  &  -1.03e-01   \\
	\hline
	\rowcolor{white}
	C$_{01}$  & -2.83e-01  & 1.94e-01 & -3.93e-01 &  8.55e-02  \\
	\hline
	\rowcolor{lightgray}
	C$_{02}$  & 2.74e-02  & -2.63e-02 &  4.47e-02  &  -1.57e-02\\
	\hline
	\rowcolor{white}
	C$_{10}$  & 9.03e-04   & 5.34e-04  &  1.49e-03  &  9.29e-04\\
	\hline
	\rowcolor{lightgray}
	C$_{11}$  &  -1.41e-04  & -2.29e-04 & -6.49e-04  &  -4.08e-04\\
	\hline
	\rowcolor{white}
	C$_{12}$  & -1.83e-05   & 2.40e-05 & 7.17e-05  &  4.73e-05\\
	\hline
	\rowcolor{lightgray}
	C$_{20}$  & 3.94e-07   & -4.47e-08 &  -1.72e-07  &  -1.52e-07\\
	\hline
	\rowcolor{white}
	C$_{21}$  &  -2.64e-07   & 2.34e-08 &  8.35e-08  &  7.55e-08\\
	\hline
	\rowcolor{lightgray}
	C$_{22}$  & 4.11e-08    & -2.85e-09 &  -1.01e-08  &  -9.65e-09\\
    \hline
    \end{tabularx}
     \end{center}
    Table 6: The polynomial coefficients for the empirical CTE-correction model when using FLT data. To apply the correction model, match the total background in the image being corrected (sky+dark+post-flash) to the closest background level in one of the tables. Use the coefficients in a given column and input them into Equation \ref{eq2}.
    \end{table}

\begin{table}[h!]
 \begin{center}
 \caption{Empirical CTE model coefficients for non-CTE-corrected FLT data \label{tab:7}}
    \vspace{.15cm}
    \begin{tabularx}{\linewidth}{|X|c|c|c|c|}
    \hline \hline  
	Filter, Exposure Time & F502N, Long & F502N, Long & F502N, Long & F502N, Long \\
	 Background [e$^-$/pix]  & $\sim$ 13-15  & $\sim$ 20-25  & $\sim$ 30-35  & $\sim$ 40-45   \\
    \hline 
    \rowcolor{lightgray}
	C$_{00}$  & 9.24e-02 &  5.30e-01 &  -1.46e-01  &  3.76e-01   \\
	\hline
	\rowcolor{white}
	C$_{01}$  & -2.67e-02  & -2.45e-01 &  2.28e-02  &  -1.79e-01 \\
	\hline
	\rowcolor{lightgray}
	C$_{02}$  & 1.34e-04  & 3.21e-02 &  5.76e-03  & 2.06e-02 \\
	\hline
	\rowcolor{white}
	C$_{10}$  & 1.46e-04  & -2.30e-04  &  5.66e-05  &  -3.57e-04 \\
	\hline
	\rowcolor{lightgray}
	C$_{11}$  & -1.60e-05  & 1.61e-04 & 5.53e-05  &  2.17e-04 \\
	\hline
	\rowcolor{white}
	C$_{12}$  & -2.05e-06  & -2.60e-05 &  -1.66e-05 &  -2.96e-05 \\
	\hline
	\rowcolor{lightgray}
	C$_{20}$  & 1.62e-08  & 6.49e-08 & 2.97e-08  &  8.13e-08 \\
	\hline
	\rowcolor{white}
	C$_{21}$  & -1.06e-08  & -3.47e-08 &  -2.26e-08  & -4.23e-08 \\
	\hline
	\rowcolor{lightgray}
	C$_{22}$  & 1.37e-09   & 4.83e-09 &  3.85e-09  &  5.34e-09 \\
    \hline
    \end{tabularx}
     \end{center}
     Table 7: The polynomial coefficients for the empirical CTE-correction model when using FLT data. To apply the correction model, match the total background in the image being corrected (sky+dark+post-flash) to the closest background level in one of the tables. Use the coefficients in a given column and input them into Equation \ref{eq2}.
    \end{table}
\begin{table}[h!]
 \begin{center}
 \caption{Empirical CTE model coefficients for CTE-corrected FLC data \label{tab:8}}
    \vspace{.15cm}
    \begin{tabularx}{\linewidth}{|X|c|c|c|c|}
    \hline \hline  
	Filter, Exposure Time & F502N, Short & F502N, Short & F502N, Long & F502N, Long \\
	 Background [e$^-$/pix]  & $\sim$ 0.1-0.5  & $\sim$ 12-12.5  & $\sim$ 1-3  & $\sim$ 7-10  \\
    \hline 
    \rowcolor{lightgray}
	C$_{00}$  & 4.51e-01  &  1.26e+00 &  1.04e+00  &  -1.33e-02   \\
	\hline
	\rowcolor{white}
	C$_{01}$  & -1.45e-01  & -7.14e-01 & -4.93e-01 &  2.86e-02  \\
	\hline
	\rowcolor{lightgray}
	C$_{02}$  & 1.00e-02  & 9.70e-02 &  5.78e-02  &  -7.48e-03\\
	\hline
	\rowcolor{white}
	C$_{10}$  & 1.50e-03   & -2.03e-04  &  -2.11e-04  &  8.80e-04\\
	\hline
	\rowcolor{lightgray}
	C$_{11}$  &  -7.58e-04  & 1.76e-04 & 1.43e-04  &  -4.20e-04\\
	\hline
	\rowcolor{white}
	C$_{12}$  & 9.40e-05   & -2.95e-05 & -2.13e-05  &  5.14e-05\\
	\hline
	\rowcolor{lightgray}
	C$_{20}$  & -1.28e-07   & 6.40e-08 &  5.88e-08  &  -1.12e-07\\
	\hline
	\rowcolor{white}
	C$_{21}$  &  6.04e-08   & -4.13e-08 &  -3.74e-08  &  5.46e-08\\
	\hline
	\rowcolor{lightgray}
	C$_{22}$  & -6.92e-09    & 6.07e-09 &  5.44e-09  &  -6.91e-09\\
    \hline
    \end{tabularx}
     \end{center}
    Table 8: The polynomial coefficients for the empirical CTE-correction model when using FLC data. To apply the correction model, match the total background in the image being corrected (sky+dark+post-flash) to the closest background level in one of the tables. Use the coefficients in a given column and input them into Equation \ref{eq2}.
    \end{table}

\begin{table}[h!]
 \begin{center}
 \caption{Empirical CTE model coefficients for CTE-corrected FLC data \label{tab:9}}
    \vspace{.15cm}
    \begin{tabularx}{\linewidth}{|X|c|c|c|c|}
    \hline \hline  
	Filter, Exposure Time & F502N, Long & F502N, Long & F502N, Long & F502N, Long \\
	 Background [e$^-$/pix]  & $\sim$ 13-15  & $\sim$ 20-25  & $\sim$ 30-35  & $\sim$ 40-45   \\
    \hline 
    \rowcolor{lightgray}
	C$_{00}$  & 2.13e-01  &  1.24e+00 &  -2.29e-01  &  2.18e-01   \\
	\hline
	\rowcolor{white}
	C$_{01}$  & -1.05e-01   & -5.78e-01 &  1.03e-01  &  -1.10e-01 \\
	\hline
	\rowcolor{lightgray}
	C$_{02}$  & 1.16e-02   & 6.92e-02 &  -1.02e-02  & 1.30e-02 \\
	\hline
	\rowcolor{white}
	C$_{10}$  & 2.17e-04   & -4.34e-04  &  4.22e-04  &  9.12e-05 \\
	\hline
	\rowcolor{lightgray}
	C$_{11}$  & -8.13e-05   & 2.08e-04 & -1.88e-04  &  -3.70e-05 \\
	\hline
	\rowcolor{white}
	C$_{12}$  & 8.02e-06   & -2.62e-05 &  2.01e-05 &  3.98e-06 \\
	\hline
	\rowcolor{lightgray}
	C$_{20}$  & 3.73e-08   & 1.13e-07 & -1.23e-08  &  2.57e-08 \\
	\hline
	\rowcolor{white}
	C$_{21}$  & -2.16e-08   & -5.51e-08 &  2.52e-09  & -1.45e-08 \\
	\hline
	\rowcolor{lightgray}
	C$_{22}$  & 2.88e-09    & 6.92e-09 &  1.92e-10  &  1.96e-09 \\
    \hline
    \end{tabularx}
     \end{center}
     Table 9: The polynomial coefficients for the empirical CTE-correction model when using FLC data. To apply the correction model, match the total background in the image being corrected (sky+dark+post-flash) to the closest background level in one of the tables. Use the coefficients in a given column and input them into Equation \ref{eq2}. \\ 
     \\
     Jump Back to the \hyperlink{page.2}{Intro}, \hyperlink{page.3}{Data}, \hyperlink{page.4}{Results}, or \hyperlink{page.8}{Conclusion} Section
    \end{table}

\end{document}